\documentclass{ws-p9-75x6-50}

\newcommand{\lesssim}{\buildrel < \over {\hspace{-.1em} {}_{\sim}} }

\newcommand{\tr}{{\mathrm{Tr}}}
\newcommand{\dett}{{\mathrm{det}}}

\newcommand{\one}{\hbox{ 1\kern-.8mm l}}

\newcommand{\bear}{\begin{array}{l}}
\newcommand{\eear}{\end{array}}
\newcommand{\ds}{\displaystyle}

\newcommand{\inte}{\! \int \!\!}
\newcommand{\infu}{\int \!\! {\cal D}[V] \prod_{i=1}^{3} {\cal D}[\varphi_i]
\, {\cal D} [\bar{\varphi}_i]}

\newcommand{\mdm}{M \partial_{M}}

\newcommand{\ie}{{\it i.e.}\ }
\newcommand{\eg}{{\it e.g.}\ }

\newcommand{\lla}{\left \langle} 
\newcommand{\rra}{\right \rangle}
\newcommand{\ug}{\!=\!}
\def\eq#1{eq.~(\ref{#1})}
\def\ceq#1{Eq.~(\ref{#1})}
\def\dphi#1#2#3{\frac{\delta #1}{\delta \varphi_{#2}(#3)} }
\def\ddphi#1#2#3#4#5{\frac{\delta^2 #1}{\delta \varphi_{#2}(#3) \, \delta
\varphi_{#4}(#5)} }
\def\dq#1#2#3{\frac{\delta #1}{\delta Q_{#2}(#3)} }
\def\dqt#1#2#3{\frac{\delta #1}{\delta \tilde{Q}_{#2}(#3)} }
\def\ddq#1#2#3#4#5{\frac{\delta^2 #1}{\delta \tilde{Q}_{#2}(#3) \, \delta
Q_{#4}(#5)} }
\def\hepth#1 {{\tt  hep-th/{#1}}}
\def\heplat#1 {{\tt  hep-lat/{#1}}}
\def\adp#1#2#3   
		{{\it Adv.\ Phys.\ }{\bf #1} (#2) #3}
\def\ap#1#2#3    
		{{\it Ann.\ Phys.\ (NY) }{\bf #1} (#2) #3}
\def\app#1#2#3   
		{{\it Astropart.\ Phys.\ }{\bf #1} (#2) #3}
\def\appol#1#2#3 
		{{\it Acta Phys.\ Polon.\ }{\bf #1} (#2) #3}
\def\arnps#1#2#3 
		{{\it Ann.\ Rev.\ Nucl.\ Part.\ Sci.\ }{\bf #1} (#2) #3}
\def\atmp#1#2#3 
		{{\it Adv.\ Theor.\ Math.\ Phys.\ }{\bf #1} (#2) #3}
\def\cpc#1#2#3   
		{{\it Comput.\ Phys.\ Commun.\ }{\bf #1} (#2) #3}
\def\cmp#1#2#3   
		{{\it Comm.\ Math.\ Phys.\ }{\bf #1} (#2) #3}
\def\dmj#1#2#3   
		{{\it Duke Math.\ J. }{\bf #1} (#2) #3}
\def\epjc#1#2#3  
		{{\it Eur.\ Phys.\ J. }{\bf C #1} (#2) #3}
\def\jmp#1#2#3   
		{{\it J.\ Math.\ Phys.\ }{\bf #1} (#2) #3}
\def\jgp#1#2#3   
		{{\it J.\ Geom.\ Phys.\ }{\bf #1} (#2) #3}
\def\jphg#1#2#3   
		{{\it J. Phys.\ }{\bf G #1} (#2) #3}
\def\cqg#1#2#3   
		{{\it Class.\ and Quant.\ Grav.\ }{\bf #1} (#2) #3}
\def\hpa#1#2#3   
		{{\it Helv.\ Phys.\ Acta }{\bf #1} (#2) #3}
\def\jhep#1#2#3 {{\it J. High Energy Phys.\ }{\bf #1} (#2) #3}
\def\lmp#1#2#3   
		{{\it Lett.\ Math.\ Phys.\ }{\bf #1} (#2) #3}
\def\npa#1#2#3   
		{{\it Nucl.\ Phys.\ }{\bf A #1} (#2) #3}
\def\npb#1#2#3    
		{{\it Nucl.\ Phys.\ }{\bf B #1} (#2) #3}
\def\npps#1#2#3  
		{{\it Nucl.\ Phys.\ }{\bf #1} {\it(Proc.\ Suppl.)} (#2) #3}
\def\pla#1#2#3   
		{{\it Phys.\ Lett.\ }{\bf A #1} (#2) #3}
\def\plb#1#2#3   
		{{\it Phys.\ Lett.\ }{\bf B #1} (#2) #3}
\def\ppnp#1#2#3  
		{{\it Prog.\ Part.\ Nucl.\ Phys.\ }{\bf #1} (#2) #3}
\def\pr#1#2#3    
		{{\it Phys.\ Rev.\ }{\bf #1} (#2) #3}
\def\pra#1#2#3   
		{{\it Phys.\ Rev.\ }{\bf A #1} (#2) #3}
\def\prb#1#2#3   
		{{\it Phys.\ Rev.\ }{\bf B #1} (#2) #3}
\def\prc#1#2#3   
		{{\it Phys.\ Rev.\ }{\bf C #1} (#2) #3}
\def\prd #1#2#3  
		{{\it Phys.\ Rev.\ }{\bf D #1} (#2) #3}
\def\pre#1#2#3   
		{{\it Phys.\ Rev.\ }{\bf E #1} (#2) #3}
\def\prep#1#2#3  
		{{\it Phys.\ Rep.\ }{\bf #1} (#2) #3}
\def\prl#1#2#3   
		{{\it Phys.\ Rev.\ Lett.\ }{\bf #1} (#2) #3}
\def\ptp#1#2#3   
		{{\it Prog.\ Theor.\ Phys.\ }{\bf #1} (#2) #3}
\def\rmp#1#2#3   
		{{\it Rev.\ Mod.\ Phys.\ }{\bf #1} (#2) #3}
\def\zpc#1#2#3   
		{{\it Z.\ Physik }{\bf C #1} (#2) #3}
\def\mpl#1#2#3a  
		{{\it Mod.\ Phys.\ Lett.\ }{\bf A #1} (#2) #3}
\def\mplb#1#2#3  
		{{\it Mod.\ Phys.\ Lett.\ }{\bf B #1} (#2) #3}
\def\sjnp#1#2#3  
		{{\it Sov.\ J.\ Nucl.\ Phys.\ }{\bf #1} (#2) #3}
\def\jetp#1#2#3  
		{{\it Sov.\ Phys.\ JETP\/ }{\bf #1} (#2) #3}
\def\jetpl#1#2#3  
		{{\it Sov.\ Phys.\ JETP Lett.\ }{\bf #1} (#2) #3}
\def\zetf#1#2#3  
		{{\it Zh.\ Eksp.\ Teor.\ Fiz.\ }{\bf #1} (#2) #3}
\def\yf#1#2#3    
		{{\it Yad.\ Fiz.\ }{\bf #1} (#2) #3}
\def\nc#1#2#3    
		{{\it Nuovo Cim.\ }{\bf #1} (#2) #3}
\def\joth#1#2#3  
		{{\it J.\ Operator Theory }{\bf #1} (#2) #3}

\def\ijmpa#1#2#3 
		{{\it Int.\ J.\ Mod.\ Phys.\ }{\bf A #1} (#2) #3}
\def\ijmpb#1#2#3 
		{{\it Int.\ J.\ Mod.\ Phys.\ }{\bf B #1} (#2) #3}
\begin{document}

\title{Applications of exact renormalization group techniques to the
non-perturbative study of supersymmetric gauge field theory}

\author{S. Arnone}

\address{Department of Physics and Astronomy,\\
University of Southampton\\
Highfield, Southampton SO17 1BJ, U.K.\\
E-mail: sa@hep.phys.soton.ac.uk}

\author{S. Chiantese and K. Yoshida}

\address{Dipartimento di Fisica,\\
Universit\`a degli Studi di Roma ``La Sapienza''\\
P.le Aldo Moro, 2 - 00185 Roma, Italia\\
and\\
I.N.F.N., Sezione di Roma I\\
E-mail: stefano.chiantese@roma1.infn.it, kensuke.yoshida@roma1.infn.it}  


\maketitle

\abstracts{
Exact Renormalization Group techniques are applied to supersymmetric models
in order to get some insights into the low energy effective actions of such
theories. Starting from the ultra-violet finite mass deformed $N\ug 4$
supersymmetric Yang--Mills theory, one varies the regularising mass and
compensates for it by introducing an effective Wilsonian
action. (Polchinski's) renormalization group equation is modified in an
essential way by the presence of rescaling ({\it a.k.a.} Konishi) anomaly,
which is responsible for the beta-function. When supersymmetry is broken up
to $N\ug 1$ the form of effective actions in terms of massless fields is
quite reasonable, while in the
case of the $N\ug 2$ model we appear to have problems related to instantons.}

\section{Introduction}
An important feature of supersymmetric gauge field theories
(supersymmetric Yang--Mills (SYM), supersymmetric QCD (SQCD), etc.) 
is that one can obtain the so called ``exact results'' for certain general class of
models.\\
\noindent
One part of these exact results is the ``non-renormalization theorem'',
which allows one to severely restrict the form of the effective (Wilsonian)
action\cite{We}.

Now all such ``exact results'' are generally demonstrated by usual
renormalized perturbation techniques, assuming moreover the existence of a
finite cutoff preserving the relevant symmetries, \ie rigid supersymmetry
and gauge symmetry.

Beyond this, one adds the effect of instantons and appeals to the absence
(in the case of supersymmetric models) of perturbative corrections to it.

It is clear that as long as one relies on such techniques, one remains in
the regime of semi-classical approximations and the results obtained are,
in principle, valid in the weak coupling limit only.

There are of course more direct, truly non-perturbative methods which are
based on the straightforward evaluation of the relevant path
integrals. They are:\\
I.\phantom{I} the lattice approximation\cite{Wil};\\
II. the Exact Renormalization Group (ERG) approach\cite{Po,Bec}.\\
With these techniques, one has the possibility of altogether leaving the
semi-classical regime and obtaining truly non-perturbative (exact) results.
\noindent
Unfortunately, both methods suffer from ``technical'' problems that cause
their applications to supersymmetric (SUSY) gauge field theories (GFT) to
be rather problematic at present.
\noindent
In the lattice approximation, one has to deal with the problem of proper
lattice definition of chiral or Weyl fermions (\ie neutrinos), which makes
the realization of SUSY on the lattice difficult.
\noindent
In the ERG approach, on the other hand, one encounters the difficulty of defining
a cutoff scheme which is compatible with gauge symmetry at arbitrary finite
cutoff value.

Recently, there has been considerable progress in understanding both of
these problems, \ie chiral fermions on the lattice\cite{lat} and gauge invariance of
ERG\cite{Mor}. However, this progress  
is still far from producing simple and exact rules to
be applied to the actual calculations and demonstrations of theorems.

In the present note, we would like to propose a method which, in view of
the difficulty encountered with gauge invariance, circumvents the cutoff
problem in the ERG approach.
\noindent
In fact, our method consists in reformulating {\it \`a la} Polchinski\cite{Po} the
original ideas by Arkani-Hamed and Murayama\cite{mu1,mu2}.\\
\noindent
Just as in the case of\cite{mu1,mu2}, the  main ingredients of our methods are
the following:\\
I.\phantom{I} the existence of ultra-violet (UV) finite models with extended supersymmetry;\\
II. rescaling anomaly in $N\ug 1$ SUSY GFT (Konishi anomaly) which
influences the relevant ERG equation in an essential way\cite{Kon,noi}.\\
\noindent
We will explain these ideas in what follows below.

\section{UV finite models and their mass deformations}

The authors of\cite{mu1,mu2} have ``regularised'' $N\ug 1$ SUSY GFT as the mass
deformation of $N\ug 4$ SYM theory. Similar models (the so called $N\ug
1^{*}$ models) have been studied by several authors in different
contexts\cite{Po2}.

The model is given by the classical action $S^* = S_{N=4} + \frac{1}{2}
\inte d^4 x \, d^2  \theta \sum_{i=1}^3 M_i^0 \varphi_i^2 + h.c.$, which
reads (written in terms of $N\ug 1$
superfields and in the ``holomorphic'' representation)
\be \label{esseN4}
\bear
{\ds S^* (V,\varphi_i,\overline{\varphi}_i; g_0) = \frac{1}{16} \inte d^4 x \, d^2
\theta \frac{1}{g_0^2} W^a_{\alpha} W^{a \alpha} +
\inte dx \, d^4 \theta  \,{\Re}e  \left( \frac{2}{g_0^2}\right) t_2(A)
\sum_{i=1}^3 \overline{\varphi}_i e^V \varphi_i +}\nonumber\\[0.25cm]
{\ds + \inte d^4 x \, d^2 \theta \, {\Re}e  \left( \frac{1}{g_0^2} \right)
\sqrt{2} \, \tr \left( \varphi_i \, [\varphi_j,\varphi_k] \right)
\frac{\epsilon^{ijk}}{3!} + \frac{1}{2} \inte d^4 x \, d^2
\theta \sum_{i=1}^3 M_i^0 \varphi_i^2 + h.c. }\\
\eear
\ee

This model, just as the original $N\ug 4$ SYM theory without 
mass terms, is believed to be finite. It has been shown that, at the
perturbative level, all the UV divergences cancel
out\cite{24}. Kovacs\cite{kovacs} has analysed again this model and his
calculations bear out the older claims.

At the perturbative level, the model suffers from apparently severe
infra-red (IR) divergences, which are, however, totally absent in the 
particular choice of gauge $\alpha \ug 1$\cite{kovacs}. One may hope,
therefore, that there is no divergence at all in gauge invariant
correlation functions.

In this note, we take up the most (and perhaps unjustifiably) optimistic
point of view, \ie that the model is really finite at the non-perturbative
level and the quantum partition function 
\be \label{zeta}
Z = \infu \exp i \left[ S^* (V,\varphi_i,\overline{\varphi}_i; g_0) +
\mbox{sources} \right]
\ee
can be well defined, perhaps with the usual procedure of gauge
fixing\footnote{${\cal D}[V]$ is meant to include accompanying ghosts.}.

As for the physics represented by \eq{zeta}, while at large energy scales,
$M_i^0 < p$, the model approaches $N\ug 4$ SYM, at low energies, $p <<
M_i^0$, $N\ug 4$ supersymmetry is broken and models with heavy fields
decoupled, such as $N\ug 1$, are obtained.

In what follows, we simplify the analysis by choosing the same mass for
some of the chiral superfields while keeping the others massless
\be \label{restrcase}
\sum_{i=1}^3 M_i^0 \varphi_i^2 \, \Rightarrow \, M_0 \sum_{i=1}^{\mu}
\varphi_i^2, \quad \mu=1,2,3.
\ee
Then at low energies we expect:\\
I.\phantom{II} $\mu=3 \quad \Rightarrow$ $N\ug 1$ SYM;\\
II.\phantom{I} $\mu=2 \quad \Rightarrow$ $N\ug 2$ SYM;\\
III. $\mu=1 \quad \Rightarrow$ $N\ug 1$ SQCD with one hypermultiplet
in the adjoint representation.


\section{Analysis of mass deformed $N\ug 4$ SYM ($N\ug 1^*$ models)}
\label{sec:tre}

We study the model given by \eq{esseN4} for the restricted case,
\eq{restrcase}. We start from the path integral expression
\be \label{zetam0}
Z_{M_0} = \infu \exp i \left[ S^* (V,\varphi_i,\overline{\varphi}_i; g_0) +
\mbox{sources} \right],
\ee
where $S^*$ is the $N\ug 4$ action deformed by the addition of $\frac{1}{2}
M_0 \sum_{i=1}^{\mu} \varphi_i^2$.

As we have argued in the previous section, we assume $Z_{M_0}$ to be well
defined, free from divergences even at the non-perturbative level\cite{kovacs}.

\subsection{$\mu = 3$ case}

We give the same mass, $M_0$, to all the chiral superfields, $\left( \varphi_i
\right)_{i=1}^3$. 
At low energies, $p<<M_0$, the model must approach $N\ug 1$ SYM and $M_0$
can be identified as the UV cutoff of such theory.\\[0.25cm]
\noindent
{\it ERG analysis}\\[0.25cm]
\noindent
As in the conventional ERG analysis\cite{Po} with momentum cutoff $\Lambda$,
one tries to vary the mass parameter, $M_0$, to a lower value, $M$, while
keeping physics unchanged\cite{mu2}. More precisely, we look for  the
effective ``Wilsonian'' action $S_M$ such that 
\be \label{zetam0eqzetam}
\bear
{\ds Z_{M_0} = Z_M \equiv \infu \exp i \bigg[ S_M
(V,\varphi_i,\overline{\varphi}_i) + 
 \frac{M}{2} \inte d^2 \theta \sum_{i=1}^{3} \varphi_i^2 +
h.c. +}\nonumber\\[0.25cm]  
{\ds + \inte d^2 \theta \sum_{i=1}^{3} f(M) \, J_i \, \varphi_i + h.c. + 
\frac{1}{2} \inte d^2 \theta \sum_{i=1}^{3} g(M) \, J_i^2 + h.c. + \inte d^4
\theta \, J_V V \bigg]. }
\eear  
 \ee
\ceq{zetam0eqzetam} is the rewriting of 
\eq{zetam0} where the change in
the mass parameter, $M_0 \rightarrow M$, is being compensated for by the
change in the action, $S_{N=4} \rightarrow S_M$, as well as by the
renormalization of external sources\footnote{We have anticipated the
necessity for a contact term.} represented by the functions $f(M)$ and
$g(M)$.\\
Just as in the case of the usual ERG approach, $S_M$ is expected to be
non-renormalizable and, in general, not even local.

To find the equation satisfied by $S_M$, one might proceed as in\cite{noi},
where the original method by Polchinski\cite{Po} is closely
followed. However, here we would like to apply the ``field redefinition''
(FRD) approach proposed by Morris\cite{Mor2,LM}. This method allows us to treat
from a unified point of view the wider class of ERG equations we want to
deal with later on.

Let us write down the generating functional, \eq{zetam0eqzetam}, as $Z_M =
\infu \exp i S_M^{tot}$ and split $S_M^{tot}$ into $S_M^0 + S_M^1$, with
the former being the mass deformation, $\frac{M}{2} \inte d^2 \theta
\sum_{i=1}^{3} \varphi_i^2 + h.c.$, and the latter standing for $S_M$ and the
renormalized source terms as well (cf \eq{zetam0eqzetam}).

We introduce the RG flow generating function (the ``RG kernel'' according
to Morris\cite{Mor2}),
\be \label{kernel}
\Psi_i(x;M) = \frac{1}{2 M^2} \dphi{}{i}{x} \left( S_M^1 - S_M^0 \right),
 \ee
and the corresponding infinitesimal change of variables 
\be \label{cov}
\varphi_i \rightarrow \varphi_i + \delta \varphi_i \equiv \varphi_i +
\delta M \, \Psi_i(x;M).
\ee
Under the FRD transformation, \eq{cov}, $Z_M$ must remain invariant
\be \label{deltapsi}
0 = \delta_{\Psi} Z_M = \delta M \!\!\infu \! \left\{ \inte \left( \dphi{\Psi_i(x)}{i}{x} + i
\Psi_i(x) \dphi{S_M^{tot}}{i}{x} \right) + h.c. \right\} e^{i S_M^{tot}}.
\ee
In the r.h.s. of \eq{deltapsi}, the first term is the Jacobian for the
functional measure under the transformation \eq{cov} while the second term
is the variation of $S_M^{tot}$. 

One can write \eq{deltapsi} in short hand as
\be \label{sh}
\left \langle \inte \left( \dphi{\Psi_i(x)}{i}{x} + i
\Psi_i(x) \dphi{S_M^{tot}}{i}{x} \right) + h.c. \right \rangle = 0.
\ee
The above relations are trivial, however \eq{sh} implies that if
$S_M^{tot}$ varies with $M$ according to 
\be \label{fleqweak}
\left \langle \mdm S_M^{tot} - M \inte \left( \dphi{\Psi_i(x)}{i}{x} + i
\Psi_i(x) \dphi{S_M^{tot}}{i}{x} \right) - h.c. \right \rangle = 0,
\ee
then $Z_M$ stays unchanged under the infinitesimal change $S_M \rightarrow
S_{M+\delta M}$.\\

\ceq{fleqweak} corresponds to the weak form of Polchinski's equation\cite{Po}.

One might ask here whether it is possible to take away the functional
average sign $\langle \cdot \rangle$ and satisfy the strong form of
Polchinski's equation
\be \label{fleqstrong}
\mdm S_M^{tot} = M \inte \left( \dphi{\Psi_i(x)}{i}{x} + i
\Psi_i(x) \dphi{S_M^{tot}}{i}{x} \right) + h.c. 
\ee
We will comment on this later on and, for the moment, carry on with the
original equation (\ref{fleqweak}).
Substituting \eq{kernel} into \eq{fleqweak} yields\footnote{Following
Polchinski, we have gone over to the momentum representation}
\be \label{fleqweakdue}
\bear
{\ds \bigg \langle \mdm S_M - \frac{1}{2 M} \inte \left( i
\ddphi{S_M}{i}{p}{i}{-p} - \dphi{S_M}{i}{p} \dphi{S_M}{i}{-p} - i M
\dphi{\varphi_i(p)}{i}{p} \right) - h.c. + }\nonumber\\[0.25cm] 
{\ds +\inte \left( f-\mdm f \right) J_i(p) \, \varphi_i(-p) + \frac{1}{2 M}
\inte \left( f^2 - M^2 \partial_M g \right) J_i(p) \, J_i(-p) + h.c. \bigg
\rangle = 0.}
\eear
\ee
If the source term renormalization factors satisfy the differential
equations $\mdm f = f$ and $M^2 \partial_M g = f^2$ with the initial
conditions $f(M_0) =1$ and $g(M_0) = 0$, then \eq{fleqweakdue} takes the
form
\be \label{fleqweaktre}
\left \langle \mdm S_M - \frac{1}{2 M} \inte \left( i
\ddphi{S_M}{i}{p}{i}{-p} - \dphi{S_M}{i}{p} \dphi{S_M}{i}{-p} - i M
\dphi{\varphi_i(p)}{i}{p} \right) - h.c. \right \rangle = 0. 
\ee
It is easy to solve the differential equations for $f$ and $g$, the
solutions corresponding to the given initial conditions being $f(M) =
\frac{M}{M_0}$ and $g(M) = \frac{1}{M_0} \left(\frac{M}{M_0} - 1 \right)$.

\ceq{fleqweaktre} has the same form as Polchinski's except for the presence
of the singular term $\Delta = -{i \over 2} \left \langle \inte
\dphi{\varphi_i(p)}{i}{p} \right \rangle$. Such a term, discarded in
Polchinski's original work, is of great importance when dealing with SUSY
GFT, as we pointed out in\cite{noi}.

Indeed, $\Delta$ is closely related to the anomalous Jacobian
(Fujikawa-Konishi determinant) under the rescaling transformation
$\varphi_i = \exp [\delta \alpha] \, \varphi'_i$, where $\delta \alpha$ is
a chiral superfield as well.
\be
J(\delta \alpha) = \dett \left( \frac{\delta \varphi_i}{\delta \varphi'_i}
\right) = 1 + i \inte \delta \alpha \frac{1}{8} \frac{t_2(A)}{8 \pi^2}
W_\alpha^a W^{a \alpha} + {\cal O}(\delta \alpha)^2,
\ee  
where $t_2(A)$ is the Dynkin index of the adjoint representation of the
gauge group and for $SU(N_C)$ one has $t_2(A) = N_C$.
Then, as we have shown in\cite{noi},
\be \label{delta}
\lla \Delta \rra = \frac{1}{16} \frac{t_2(A)}{8 \pi^2}
\inte W_\alpha^a W^{a \alpha}  
\ee


Substituting \eq{delta} into \eq{fleqweaktre} after summing upon $i$ and
assuming 
the strong form for the rest of the equation, \ie taking away the
functional average sign, one gets
\be \label{fleqweakquattro}
\mdm S_M = \frac{1}{2 M} \inte \sum_{i=1}^3 \left( i
\ddphi{S_M}{i}{p}{i}{-p} - \dphi{S_M}{i}{p} \dphi{S_M}{i}{-p} \right) +
\frac{3}{16} \frac{t_2(A)}{8 \pi^2} \inte  W^2 + h.c., 
\ee
which is the main result in\cite{noi}.

One can explicitly separate from $S_M$ the contribution of the anomalous
term by introducing the ``normal'' part of the action, $\tilde{S}_M$, 
\be \label{smtilde}
\tilde{S}_M  \doteq
S_M - \frac{3}{16} \frac{t_2(A)}{8 \pi^2} \inte  W^2 \log \left(
\frac{M}{M_0} \right) +h.c.
\ee
It satisfies Polchinski's equation (without the anomalous term $\Delta$)
\be \label{nonanomeq}
\mdm \tilde{S}_M = \frac{1}{2 M} \inte \sum_{i=1}^3 \left( i
\ddphi{\tilde{S}_M}{i}{p}{i}{-p} - \dphi{\tilde{S}_M}{i}{p} \dphi{\tilde{S}_M}{i}{-p} \right) 
+ h.c. 
\ee
\noindent
{\it The variation of the gauge coupling constant}\\[0.25cm]
\noindent
As it has been argued in\cite{noi}, one expects the contribution of the
anomalous part to be dominant at $M \neq M_0$. In fact $\tilde{S}_M \sim  \tilde{S}_{M_0}
+{\cal O} \left( \frac{1}{M},\frac{1}{M_0} \right)$. Thus, at low energies,
one may read off the variation of the gauge coupling constant from
$\Delta$, \eq{delta}
\be \label{changeing}
\frac{1}{g^2(M)} - \frac{1}{g^2(M_0)} = \frac{3 \, t_2(A)}{8 \pi^2} \log \left(
\frac{M}{M_0} \right).    
\ee
This is the variation of the ``holomorphic'' gauge coupling constant. One
can go over to the more conventional canonical representation by 
rescaling all the superfields by the canonical gauge coupling, \ie
$\varphi_{(h)} = g_c \, \varphi_{(c)}$, $\overline{\varphi}_{(h)} =
g_c \, \overline{\varphi}_{(c)}$ and $V_{(h)} = g_c \, V_{(c)}$, and, of
course, taking into account the effect of Konishi anomaly in such a
transformation. We refer to\cite{mu2} for the details and simply quote the result
\be \label{SV}
\beta(g_c) = \mdm g_c(M) = - \frac{\ds \frac{3}{16 \pi^2} \, t_2(A) \,
g_c^3}{\ds 1 - \frac{t_2(A)}{8 \pi^2} \, g_c^2}.
\ee
\ceq{SV} is the celebrated Novikov-Shiffman-Vainshtein-Zakarov (NSVZ)
exact expression of the $\beta$-function in $N\ug 1$ SYM.\\[0.25cm]
\noindent
{\it The infinite-$M_0$ limit}\\[0.25cm]
\noindent 
The conventional infinite-$M_0$ limit is defined by the conditions $M_0 \rightarrow
\infty$ and $g(M_0) \rightarrow 0$ in such a way as to keep the dynamical
cutoff $\Lambda = M_0 \exp - \left(\frac{8 \pi^2}{g_0^2} \frac{1}{3
N_C}\right)$ fixed. According to \eq{changeing}, $\Lambda$ is the value of
$M$ at which $g(M)$ diverges.\\
\noindent
Then, again with  $\tilde{S}_M \sim
\tilde{S}_{M_0}$, one gets
\be
S(M) \sim \frac{1}{16} \inte \left[ \frac{1}{g_0^2} + \frac{3 N_C}{8 \pi^2} \log 
\frac{M}{M_0} \right] W^2 + h.c. = \frac{1}{16} \frac{3 N_C}{8 \pi^2} \inte \log 
\left(\frac{M}{\Lambda} \right) W^2 + h.c.
\ee
In this limit, the theory is a pure $N\ug 1$ SYM.

\subsection{$\mu = 2$ case}

One can trivially extend the previous analysis to the case when only two of
the three chiral superfields in the $N\ug 4$ SYS get massive, \ie one adds
the mass term $S^0_{M_0} = \frac{M_0}{2} \inte d^2 \theta
\sum_{i=1}^{2} \varphi_i^2 + h.c.$\\
\noindent
Note that $S^0_{M_0}$ is $N\ug 2$ SUSY invariant, thus one expects that the
original $N\ug 4$ SYM gets broken up to a $N\ug 2$ model ($N\ug 2$ SYM).

Repeating the method explained above, one obtains the
analog of \eq{smtilde} for the effective
action as
\be \label{smtildendue}
\tilde{S}_M  \doteq
S_M - \frac{2}{16} \frac{t_2(A)}{8 \pi^2} \inte  W^2 \log \left(
\frac{M}{M_0} \right) + h.c.,
\ee
where $\tilde{S}_M$ again satisfies Polchinski's equation with respect to
the heavy fields $\varphi_{1,2}$.

The variation of the gauge coupling constant can be obtained in the same
way as before; the result for the canonical $\beta$-function is
\be
\beta(g_c) = \mdm g_c(M) = - \frac{2}{16 \pi^2} \, t_2(A) \,
g_c^3,
\ee
this representing the well known fact that the $\beta$-function of $N\ug 2$
SYM is one-loop exact.

In the infinite-$M_0$ limit with fixed $\Lambda = M_0 \exp - \left(\frac{8
\pi^2}{g_0^2} \frac{1}{2 N_C}\right)$
\be \label{essendue}
S(M) \sim \frac{1}{16} \frac{2 N_C}{8 \pi^2} \inte \log 
\left(\frac{M}{\Lambda} \right) W^2 + h.c. + \varphi_3 \, \mbox{kinematical term}.
\ee
The coefficient of such a kinematical term is completely determined by
$N\ug 2$ SYM\cite{mu2}.

\subsection{$\mu = 1$ case}

Let us give mass to just one of the three chiral superfields in the $N\ug
4$ SYM, say $\varphi_1$. Then 
the mass term reads $S^0_{M_0} = \frac{M_0}{2} \inte d^2 \theta
\varphi_1^2 + h.c.$\\
\noindent
The effective action is defined by
\be \label{smtildenew}
S_M = \tilde{S}_M + \frac{1}{16} \frac{t_2(A)}{8 \pi^2} \inte  W^2 \log \left(
\frac{M}{M_0} \right) + h.c.,
\ee
where $\tilde{S}_M$ is now contributed by the chiral kinematical terms
$\sum_{i=2}^3 \overline{\varphi}_i e^V \varphi_i$.
In contrast with the previous $\mu = 2$ case, the coefficients of such
terms at arbitrary mass $M$ are no more constrained by SUSY and, thus, one must write them as 
$Z(M,M_0) \sum_{i=2}^3 \overline{\varphi}_i e^V \varphi_i$.

Rescaling superfields so as to cancel this $Z$ factor yields an extra
contribution to the $W^2$ term
\be \label{changeingnew}
\frac{1}{g^2(M)} - \frac{1}{g^2(M_0)} = \frac{t_2(A)}{8 \pi^2} \left[ \log 
\frac{M}{M_0} - 2 Z(M,M_0) \right].    
\ee

\section{Generalised RG flow} \label{sec:quattro}

In sec.~\ref{sec:tre} we have seen that the RG flow with respect to the
variation of the mass parameter is generated by the continuous FRD $\delta
\varphi_i = \delta M \, \Psi_i(x;M)$, where the generating function
$\Psi_i$ is defined in \eq{kernel}.\\ 
\noindent
We have chosen $S_M^0$ to be the mass term in $S_M^{tot}$; however, in the
case of $\mu = 1,2$ one may try a more general form of $S_M^0$ - or of
$S_{M_0}^0$ which, because of the flow equation, determines $S_M^0$ uniquely.

Note, first of all, the fact that in these cases the characteristic
trilinear term in the $N\ug 4$ action is either linear or quadratic in the 
massive chiral superfields. Thus, for the $\mu = 1,2$ cases, it is natural
to try using the following ``quadratic'' actions as $S_{M_0}^0$\\
I.\phantom{I} $\mu \ug 1 \Rightarrow S_{M_0}^0 \ug \frac{M_0}{2} \inte d^2 \theta
\varphi_1^2 + \inte \left( \vec{J}_1 + i \alpha \vec{\varphi}_2 \wedge
\vec{\varphi}_3 \right) \cdot \vec{\varphi}_1 + h.c.$, $\, \alpha
\doteq {\Re}e  \left( \frac{2 \sqrt{2}}{g_0^2} \right) t_2(A)$;\\[0.17cm] 
II. $\mu \ug 2 \Rightarrow S_{M_0}^0 \ug \frac{M_0}{2} \inte d^2 \theta
\sum_{i=1}^2 \varphi_i^2 + i \alpha \inte \left( \vec{\varphi}_1 \wedge
\vec{\varphi}_2 \right) \cdot \vec{\varphi}_3 + \sum_{i=1}^2 J_i \,
\varphi_i + h.c.$\\

To see whether these quadratic actions generate a consistent RG flow, let
us consider a generic $S_M^0$ with both quadratic and linear couplings
\be \label{quardaction}
S_M^0 = \frac{1}{2} \inte \varphi_i \, {\cal M}_{ij} (M) \, \varphi_j + h.c. +    
\inte 
J_i \, {\cal F}_{ij} (M) \, \varphi_j + h.c. + \inte 
J_i \, {\cal G}_{ij} (M) 
\, J_j + h.c.,
\ee
where the matrices ${\cal M}, {\cal F}$ and ${\cal G}$ can depend on the
physical massless fields, \ie $\left(V,\varphi_k, \overline{\varphi}_k
\right)_{\mu +1 \leq k \leq 3}$.\\ 
\noindent  
The RG flow is now generated by the FRD $\delta \varphi_i = \delta M \,
\Psi_i$ with\footnote{As $S_M^0$ now contains the renormalized source terms  
too, $S_M^1$ is just the Wilsonian action, $S_M$.}
\be
\Psi_i(x;M) =  \frac{1}{2 M^2} \dphi{}{i}{x} \left( S_M - S_M^0 \right) = 
\frac{1}{2 M^2} \left( \dphi{S_M}{i}{x} - {\cal F}_{ij} \, J_j - {\cal
M}_{ij} \, \varphi_j \right).
\ee

Applying the above FRD to the generating functional yields 
\be \label{genfleqweak}
\bear
{\ds \Bigg \langle \mdm S_M - \frac{1}{2 M} \inte \left( i
\ddphi{S_M}{i}{x}{i}{x} - \dphi{S_M}{i}{x} \dphi{S_M}{i}{x} - i {\cal
M}_{ij} \, 
\dphi{\varphi_j(x)}{i}{x} \right) - h.c. + }\nonumber\\[0.25cm] 
{\ds +\frac{1}{2} \inte \left( \frac{1}{M} \varphi^T {\cal M}^2 \varphi - 
\varphi^T \mdm {\cal M} \, \varphi \right) + h.c. +
\inte \left( \frac{1}{M} J^T {\cal F M} \, \varphi - 
J^T \mdm {\cal F} \, \varphi \right) + }\nonumber\\[0.25cm]
{\ds + \frac{1}{2}
\inte \left( \frac{1}{M} J^T {\cal F}^2 J - 
J^T \mdm {\cal G} \, J \right)+ h.c. \Bigg
\rangle = 0.}
\eear
\ee

One sees that $S_M$ satisfies Polchinski's equation with the anomalous term 
provided that the $M$-dependent matrices ${\cal M}, {\cal F}$ and ${\cal
G}$ are chosen in such a way as to satisfy
\be \label{matrixde}
\mdm {\cal M} = \frac{1}{M} {\cal M}^2, \qquad \mdm {\cal F} = \frac{1}{M}
{\cal F M}, \qquad \mdm {\cal G} = \frac{1}{M} {\cal F}^2.
\ee
Rewriting \eq{genfleqweak} in the strong form
\be \label{genfleqstrong}
\mdm S_M = \frac{1}{2 M} \inte \left( i
\ddphi{S_M}{i}{x}{i}{x} - \dphi{S_M}{i}{x} \dphi{S_M}{i}{x} - i {\cal
M}_{ij} \, 
\dphi{\varphi_j(x)}{i}{x} \right) + h.c.  
\ee
\ceq{genfleqstrong} has the same form as \eq{fleqweakquattro} except for
the anomalous term which, however, in the cases of interest can be reduced  
to the previous case. The important difference is its initial condition:
$S_{M_0}$ does not contain the characteristic trilinear term in the $N \ug
4$ action which has been explicitly separated and included in $S_M^0$ and,
as a consequence, it is just $N\ug 1$ invariant. We will comment further on this
issue in sec.~\ref{sec:conclusions}.

The matrix differential equations, \eq{matrixde}, with the initial conditions
${\cal M}(M_0) = {\cal M}_0$, ${\cal F}(M_0) = \one$ and ${\cal G}(M_0) = 0$
can be easily integrated by noting that $\left[{\cal M} , {\cal M}_0
\right] = 0$. One gets
\be \label{sols}
{\cal M}^{-1} - {\cal M}_0^{-1} = \frac{1}{M} - \frac{1}{M_0} \doteq
\frac{1}{\tilde{M}}, \qquad {\cal F} = {\cal M}{\cal M}_0^{-1}, \qquad
{\cal G} = {\cal M}{\cal M}_0^{-2} - {\cal M}_0^{-1}.
\ee
 
These results can be applied to the specific examples listed at the
beginning of sec.~\ref{sec:quattro}. 

\subsection{$\mu = 1$ case}

This is the simpler case because the mass matrix, ${\cal M}$, is just a
number, the regularising mass $M$.\\
\noindent 
\ceq{quardaction} takes the form
\be \label{mu1}
\bear
{\ds S_{M}^0 = \frac{M}{2} \inte d^2 \theta \, 
\varphi_1^2 + h.c. + \frac{M}{M_0} \inte \left( \vec{J}_1 + i \alpha \vec{\varphi}_2 \wedge
\vec{\varphi}_3 \right) \cdot \vec{\varphi}_1 + h.c. + \frac{1}{2} \left(
\frac{M}{M_0^2} - \frac{1}{M_0} \right) \times }\nonumber\\[0.25cm]
{\ds \times \inte \left( \vec{J}_1 +
i \alpha \vec{\varphi}_2 \wedge \vec{\varphi}_3 \right)^2 + h.c.}
\eear
\ee

The total action, $S_M^{tot}$, is obtained by adding to \eq{mu1} the
effective action, $S_M$, and the (un-renormalized) source terms for massless
fields.\\
\noindent
$S_M$ should satisfy Polchinski's equation with the anomalous term, which
can be again evaluated by Konishi anomaly. It reads $\Delta = \frac{1}{16}
\frac{t_2(A)}{8 \pi^2} \inte  d^2 \theta \, W^2$. 

Getting rid of $\Delta$ by defining $\tilde{S}_M = S_M -
\frac{1}{16} \frac{t_2(A)}{8 \pi^2} \inte  W^2 \log \left(\frac{M}{M_0} \right) + h.c.$, 
one obtains the effective action for low energy physical configurations,
\ie $\vec{J}_1 = \vec{\bar{J}}_1 = \vec{0} $ and $p \lesssim M << M_0$
\be \label{noteq}
S_M \ug \frac{1}{16} \frac{t_2(A)}{8 \pi^2} \inte \log \left(\!
\frac{M}{\Lambda} \! \right) 
W^2 + \frac{\alpha^2}{2 M_0} \inte \left( \vec{\varphi}_2 \wedge
\vec{\varphi}_3 \right)^2 + h.c. + \inte {\Re}e \! \left( \!
\frac{2}{g_0^2}\! \right) t_2(A)
\sum_{i=2}^3 \overline{\varphi}_i e^V \varphi_i,
\ee
where $\Lambda = M_0 \exp - \left(\frac{8
\pi^2}{g_0^2} \frac{1}{t_2(A)}\right)$.

The notable feature of \eq{noteq} is the effective four-field
interaction. Moreover, it is not necessary to integrate out the massive
$\varphi_1$ field to get it. Another important feature of the above
equation is that, in contrast to sec.~\ref{sec:tre}, there is no way of
getting the non-trivial wave function renormalization constants of massless
$\varphi_{2,3}$.

\subsection{$\mu = 2$ case}

Giving the same mass $M_0$ to $\varphi_1$ and $\varphi_2$, one expects the
resultant theory to be $N\ug 2$ SUSY. The issue is whether one can get
some information about the low energy effective action of \eg $N\ug 2$ SYM
after the heavy fields decouple.

Note that $S_{M_0}^0$ by itself is not $N\ug 2$ invariant, as it does not
contain, by construction, the required kinetic terms for the massive chiral
superfields.

The generalised mass matrix in $S_M^0$, ${\cal M}_{ia,jb} (M)$ ($ij$ run
from $1$ to $2$ whereas $ab$ label the adjoint representation of
$SU(N_C)$), can be written down as ${\cal M} = {\cal M}_0 \left( \one -
\frac{{\cal M}_0}{\tilde{M}} \right)^{-1}$, with the initial value being
read off from $S_{M_0}^0$
\be \label{emmezero}
\left( {\cal M}_0 \right)_{ia,jb} = M_0 \delta_{ij} \, \delta_{ab} + i
\alpha \varphi^c f^{cab} \varepsilon_{ij}.
\ee
Hereafter we will refer to the massless physical field,$\varphi_3$, as
$\varphi$. Thus, in \eq{emmezero}, $\varphi^c \equiv \varphi_3^c$;
$f^{cab}$'s are $SU(N_C)$ structure constants. 

For later purpose it is convenient to introduce a new matrix, $\Phi$,
defined by $\Phi \doteq \exp \left[ {\cal M} {\cal M}_0^{-1} \right]$. 
$\Phi$ satisfies $\mdm \Phi = \frac{{\cal M}}{M}$ with $\Phi(M_0) = 0$.
  
From \eq{emmezero} one can compute the matrix element $\Phi_{ia,jb}$.\\
\noindent 
We give here the explicit result for $N_C \ug 2$
\be \label{phi}
\bear
{\ds \Phi_{ia,jb}(M) = \delta_{ij} \, \delta_{ab} \log \frac{M}{M_0} -
\frac{1}{2} \delta_{ij} \left( \delta_{ab} - \frac{\varphi^a
\varphi^b}{\varphi^2} \right) \log \left( 1 + \frac{\alpha^2
\varphi^2}{M_0^2} \left(1-\frac{M}{M_0}\right)^2\right)
+}\nonumber\\[0.25cm]
{\ds + \frac{1}{\sqrt{\varphi^2}} \varphi^c \epsilon_{abc} \varepsilon_{ij} 
\log \left( \frac{1 - i \frac{\alpha}{M_0} \sqrt{\varphi^2}
\left(1-\frac{M}{M_0}\right)}{1 + i \frac{\alpha}{M_0} \sqrt{\varphi^2}
\left(1-\frac{M}{M_0}\right)} \right)}
\eear
\ee

As before, $S_M^{tot} = S_M^0 + S_M + \mbox{unrenormalized source terms}$
and the Wilsonian effective action is again split as $S_M = \tilde{S}_M + 
S_M^{anom}$, where the latter is the anomalous contribution. $\tilde{S}_M$
satisfies the anomaly free Polchinski's equation 
\be 
\mdm \tilde{S}_M = \frac{1}{2 M} \inte \sum_{i=2}^3 \left( i
\ddphi{\tilde{S}_M}{i}{x}{i}{x} - \dphi{\tilde{S}_M}{i}{x}
\dphi{\tilde{S}_M}{i}{x} \right) + h.c.  
\ee
with the reduced initial condition, \ie the initial value $S_{M_0}$ does
not contain the trilinear coupling $\varphi_3 \cdot \varphi_1 \wedge
\varphi_2$.

As for the anomalous part, one can easily appreciate it is given by
\be
S_M^{anom} = - \frac{i}{M} \lla \inte \Phi_{ia,jb}
\frac{\delta \varphi_j^b}{\delta \varphi_i^a} \rra + h.c.
 \ee
As in the previous cases, one expects the dominance of $S_M^{anom}$ over
$\tilde{S}_M$. 

To obtain the explicit expression of the anomalous contribution, we 
work assuming the following simplifying conditions:\\
I.\phantom{I} the gauge group is $SU(2)$;\\
II. we specialise to the massless sector, \ie to the configuration which
remains massless in the classical vacuum represented by the vacuum
expectation value (vev) $\lla \varphi \rra = (0,0,a)$
\be
\vec{V} = (0,0,V) \qquad \vec{\varphi} = (0,0,\varphi) \qquad
\vec{\bar{\varphi}} = (0,0,\bar{\varphi}).
\ee

For such a configuration, the relevant matrices
${\cal M, F}$ become diagonal. In particular, from \eq{phi}
\be
\Phi = \left\{ \log \frac{M}{M_0} -
\frac{1}{2} \log \left( 1 + \frac{\alpha^2
\varphi^2}{M_0^2} \left(1-\frac{M}{M_0}\right)^2\right) \right\} \!\! \one +
\frac{1}{2} 
\log \left( \frac{1 - i \frac{\alpha \varphi}{M_0} 
\left(1-\frac{M}{M_0}\right)}{1 + i \frac{\alpha \varphi}{M_0} 
\left(1-\frac{M}{M_0}\right)} \right) \sigma_3,
\ee
where $\sigma_3 = \left( \! \begin{array}{cc}
1 & 0\\[0.07cm]
0 & -1\\
\end{array} \! \right)$.

In the massless sector, it is convenient to make heavy field variables,
$\varphi_i^a, i=1,2, a=1,2$, undergo the following linear transformation
while keeping $\varphi_i^3$ unchanged  
\be \label{phims} \left\{
\bear  
{\ds \varphi_1^1 = \frac{1}{2} \left[ \left( \tilde{Q}_1 - i \tilde{Q}_2
\right) + \left( Q_1 + i Q_2 \right) \right], \quad \, \varphi_1^2 =
\frac{1}{2i} \left[ \left( \tilde{Q}_1 - i \tilde{Q}_2 \right) - 
\left( Q_1 + i Q_2 \right) \right], }\nonumber\\[0.25cm]
{\ds \varphi_2^1 = \frac{1}{2} \left[ \left( \tilde{Q}_1 + i \tilde{Q}_2
\right) + \left( Q_1 - i Q_2 \right) \right], \quad \, \varphi_2^2 =
\frac{1}{2i} \left[ \left( \tilde{Q}_1 + i \tilde{Q}_2 \right) - 
\left( Q_1 - i Q_2 \right) \right]. }\nonumber\\
\eear \right.
\ee
In terms of the new variables, $\left( Q_i, \tilde{Q}_i \right)_{i=1}^2$,
the heavy field gauge couplings become 
\be
\sum_{i=1}^2 \bar{\varphi}_i^a \left( e^{\widehat{V}} \right)_{ab} \varphi_i^b =
\sum_{i=1}^2 \left( Q_i^{\dagger} e^{-V} Q_i + \tilde{Q}_i e^{V}
\tilde{Q}_i^{\dagger} + \bar{\varphi}_i^3 \varphi_i^3 \right), 
\ee
with $\widehat{V} = V \sigma_3$.

Thus the calculation of the anomaly reduces to 
\be \label{abanom}
- \frac{i}{2} \lla \inte \frac{\delta Q_i(p)}{\delta Q_i(p)} \rra = 
- \frac{i}{2} \lla \inte \frac{\delta \tilde{Q}_i(p)}{\delta \tilde{Q}_i(p)} \rra =
 \frac{1}{16} \frac{1}{8 \pi^2} \inte W^2.
\ee
From eqs (\ref{phims}), (\ref{abanom}) 
\be
S_M^{anom} = \frac{1}{16} \frac{1}{4 \pi^2} \left[\log \left( \frac{M}{M_0} \right)^2  
- \log \left( 1 + \frac{\alpha^2 \varphi^2}{M_0^2}
\left(1-\frac{M}{M_0}\right)^2\right) \right] W^2 + h.c. 
\ee

On the other hand, Polchinski's equation for the non-anomalous action,
$\tilde{S}_{M}$, reads 
\be 
\mdm \tilde{S}_M = \frac{1}{M} \inte \sum_{i=1}^2 \left( i
\ddq{\tilde{S}_M}{i}{p}{i}{-p} - \dqt{\tilde{S}_M}{i}{p} \dq{\tilde{S}_M}{i}{-p} \right)+ h.c.,  
\ee
with the initial condition 
\be
\tilde{S}_{M_0} = 2 \inte d^4 \theta  \,
\sum_{i=1}^2 {\cal R}e  \left( \frac{2}{g_0^2}\right)
\left(Q_i^{\dagger} \, e^{-V} Q_i + \tilde{Q}_i \, e^{V} \tilde{Q}_i^{\dagger}
\right) + \mbox{irrelevant terms}.
\ee

In the end we obtain the approximate expression of the Wilsonian effective
action
\be \label{essenduegen}
S_M = -
\frac{1}{64 \pi^2} \inte \left[\log \left( \frac{\Lambda}{M} \right)^2  
+ \log \left( 1 + \frac{\alpha^2 \varphi^2}{M_0^2}
\left(1-\frac{M}{M_0}\right)^2\right) \right] W^2 + \varphi \, \mbox{kinetic term}, 
\ee
with $\Lambda$ being $M_0 \exp - \left(\frac{2 \pi^2}{g_0^2} \right)$.

Apart from the first logarithmic term, which is the same as in
\eq{essendue}, \eq{essenduegen} is not very satisfactory\cite{SW1,SW2}. The obvious
trouble with our method is the fact that $N\ug 2$ SYM is not maintained at
an arbitrary value of $M$. It is actually broken in two different
(independent) places:\\
I.\phantom{I} $S_{M_0}^0$ is not $N\ug 2$ invariant and neither is ${\cal
M}(M;\varphi)$;\\
II. Konishi anomaly is evaluated with $N\ug 1$ but not $N\ug 2$ symmetric
regulator\cite{Kon}.

Now, $N\ug 2$ generalization of\cite{Kon} has been proposed in\cite{mu2}. If
one followed their method directly, one would end up with an effective
action of the form\cite{mu2}
\be
S_M = \frac{1}{16 \pi^2} \inte \left[ F(\varphi^2) \, W^2 + h.c. + 
\bar{\varphi} \, F(\varphi^2) \, \varphi \right], 
\ee
which is $N\ug 2$ symmetric only if $F(\varphi^2)$ is constant. As it is
well known\cite{Pes}, in fact, the general $N\ug 2$ symmetric action would be of the
form
\be
S_M \propto \inte \left( 4 \varphi^2 f'' + 2 f' \right) \, W^2 + h.c. + 
2 \inte \bar{\varphi} \left( 2 f' \right) \varphi,
\ee
with $f=f(\varphi^2)$.

\section{Conclusions}
\label{sec:conclusions}

As we have seen, by applying ERG techniques to mass deformed models, one
arrives at reasonable results when the residual supersymmetry (after heavy
fields decouple) is $N\ug 1$. This is true for both UV finite $N\ug 4$ SYM
and $N\ug 2$ SQCD.

Apart from the correct expression of the relevant beta functions (NSVZ and
generalizations), we have obtained low energy effective actions for
``massless'' fields which are consistent with known results\cite{SW1,SW2}. 

It appears that the inability to construct a manifestly $N\ug 2$ symmetric
RG flow in our context always causes troubles. One interpretation is that
the final state the system flows into after heavy fields have
decoupled, is not the pure $N\ug 2$ state but a mixture of $N\ug 1$ and
$N\ug 2$ states instead\cite{tha}.\\ 
At present, we do not know how to remedy
such a situation by, \eg, projecting the final state into pure $N\ug 2$.

It is not true, however, that the effective action in \eq{essenduegen} is
totally fictitious. Inspecting the expression and comparing it with the
correct expression\cite{SW1,SW2} one notes the absence of what can be interpreted
as instanton contributions.\\
Indeed, \eq{essenduegen} can be rewritten, in the massless sector, as
\be
S(M) = \frac{1}{16} \frac{1}{4 \pi^2} \inte \sum_{\pm} \left[ \frac{2
\pi^2}{g_0^2} - \log \left( \frac{M_0 \pm i \alpha a}{\Lambda} \right) \right],
 \ee
which is consistent with the result on $N\ug 2$ SQED in\cite{SW2} (there is
no instanton in SQED).\\
It contrasts with the derivation of NSVZ beta function, in which case
instanton effects need not be taken into account separately.

All these facts seem to indicate that though
our method is capable of revealing part of the non-perturbative structure
of low energy super gauge field theory, on the other hand there are still
some important elements lacking.

One must also mention that recently a series of remarkable
insights into this class of quantum field theory models - including a
simple derivation of Seiberg-Witten results - are getting discovered by
``stringists''. The main techniques here are brane technology and AdS/CFT
correspondence.
It is left to see whether this state of affairs really means that the super
gauge field theory is just a ``spin off'' (in the sense of E. Witten) of
super string theory, or M theory.
 


\end{document}